\documentclass[mathleft
]{an}
\usepackage{graphicx}
\usepackage{times}
\begin{document}

\Pagespan{789}{}
\Yearpublication{2008}%
\Yearsubmission{2008}%
\Month{1}%
\Volume{1}%
\Issue{1}%

\title{Interferometer observations of molecular gas in radio galaxies}

\author{S.~Garc\'{\i}a-Burillo\inst{1}\fnmsep\thanks{Corresponding author:
  \email{s.gburillo@oan.es}\newline}
\and  F.~Combes\inst{2}
\and  A.~Usero\inst{1,3}
\and  A.~Fuente\inst{1}
}
\titlerunning{Molecular gas in radio galaxies}
\authorrunning{S.~Garc\'{\i}a-Burillo et al.}
\institute{
Observatorio de Madrid, OAN, Alfonso XII, 3, E-28014 Madrid, Spain
\and 
Observatoire de Paris, LERMA, 61 Av. de l'Observatoire, F-75014 Paris,
France
\and
Centre for Astrophysics Research, University of Hertfordshire, College 
Lane, Hatfield AL10 9AB,
 United Kingdom}

\received{30 Sep 2008}
\accepted{11 Oct 2008}
\publonline{later}


\abstract{We present the first results of a high-resolution study of the distribution and kinematics of molecular gas in two nearby radio galaxies, 4C~31.04 and 3C~293, representative of two different stages of evolution in radio-loud active galactic nuclei (AGN). These observations, conducted with the IRAM Plateau de Bure Interferometer (PdBI), map with unprecedented spatial resolution ($\sim$0.5$^{\prime\prime}$--1$^{\prime\prime}$) and sensitivity the emission and absorption of key molecular species such as CO, HCN and HCO$^+$. We report on the detection of a kinematically disturbed and massive (M$_{gas}\sim$10$^{10}$M$_{\odot}$) molecular/dusty disk of $\sim$1.4~kpc-size fueling the central engine of the compact symmetric object (CSO) 4C~31.04. We also report on the detection of a massive (M$_{gas}\sim$10$^{10}$M$_{\odot}$) regularly rotating $\sim$7~kpc-size disk in the FR~II radio galaxy 3C~293. A complex system of molecular line absorptions is detected against the mm-continuum source of this galaxy (AGN and jet).  We compare the properties of the molecular disks in the two sources and discuss them in the light of the different theories describing the evolution of radio galaxies.}

\maketitle

\section{Introduction}

\begin{figure*}
\begin{center}
\includegraphics[width=0.85\textwidth]{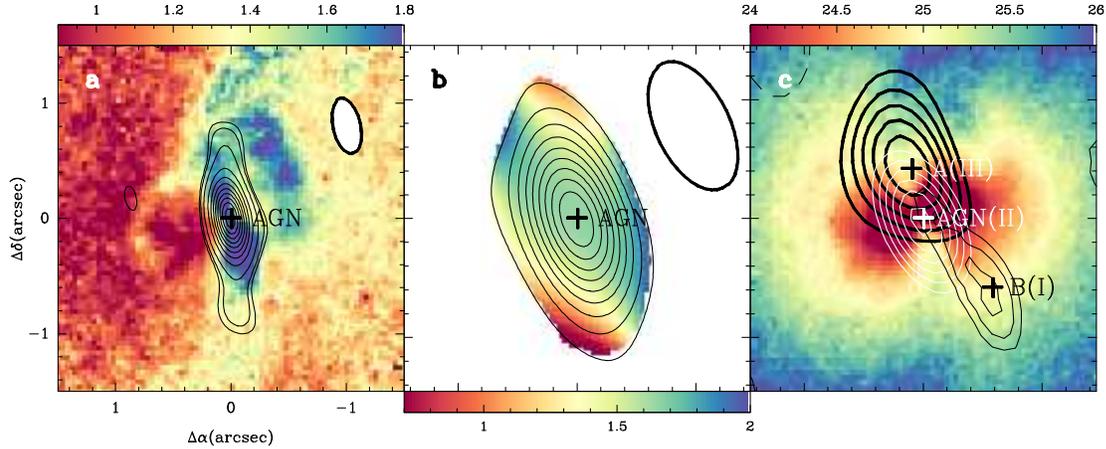}
\caption{{\bf a)} The 1mm-continuum PdBI map of 4C\,31.04 (contours: 3$\sigma$, 4$\sigma$, 7$\sigma$ to 55$\sigma$ in steps of 6$\sigma$; 1$\sigma$=0.8\,mJy\,beam$^{-1}$) is overlaid on the R-H color image of Perlman et al.~(\cite{per01}) from HST (color scale). ($\Delta\alpha$, $\Delta\delta$)--offsets are relative to the AGN locus. {\bf b)} The 1mm-continuum map, degraded to the resolution of the 3mm observations (contours: 10$\%$ to 90$\%$ in steps of 10$\%$ of the peak value), is overlaid on the spectral index map (in color scale) of the continuum emission derived between 84\,GHz and 218\,GHz. {\bf c)} The HCO$^{+}$\,(1--0) line maps of 4C\,31.04 for the three velocity channels defined in Fig.\,\ref{4c31-2}b: {\bf I}(emission)(black thin contours; with maximum at {\bf B}), {\bf II}(absorption)(white contours; with maximum at the AGN), {\bf III}(emission)(black thick contours; with maximum at {\bf A}). Levels are 3.5$\sigma$, 4$\sigma$, and 4.5$\sigma$ for channel {\bf I} (1$\sigma$({\bf I})=0.06Jy\,km\,s$^{-1}$), -40$\sigma$, to -100$\sigma$ in steps of -10$\sigma$ for channel {\bf II} (1$\sigma$({\bf II})=0.05Jy\,km\,s$^{-1}$) and 5$\sigma$ to 13$\sigma$ in steps of 2$\sigma$ for channel {\bf III} (1$\sigma$({\bf III})=0.08Jy\,km\,s$^{-1}$). HCO$^{+}$ contours are overlaid on the F702W HST image of Perlman et al.~(\cite{per01}) (color scale in mag.pixel$^{-1}$). Ellipses represent the beams at 218\,GHz({\bf a)}) and 84\,GHz({\bf b)}). See GB07 for details.}
\label{4c31-1}
\end{center}
\end{figure*}

\begin{figure*}
\begin{center}
\includegraphics[width=0.7\textwidth]{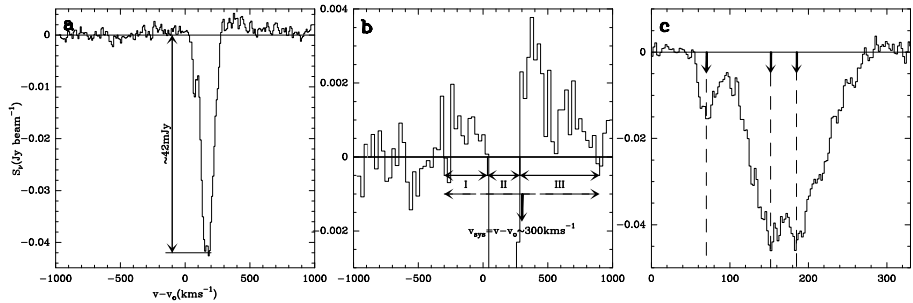}
\caption{{\bf a--b)} Spectrum of the HCO$^{+}$(1--0) line emission and absorption at the central offset (AGN) of 4C\,31.04. A strong absorption line is detected against the 3mm continuum source. Velocity scale is relative to $z$=0.0592. {\bf b)} A zoom on the HCO$^{+}$(1--0) spectrum of {\bf a)}. Intervals {\bf I}--to--{\bf III} and v$_{sys}$ ($z$=0.0602) value are highlighted. {\bf c)} A high-resolution version of the HCO$^{+}$(1--0) absorption profile of {\bf a)}; the three distinct velocity components identified in the spectrum are all blueshifted with respect to v$_{sys}$. See GB07 for details.}
\label{4c31-2}
\end{center}
\end{figure*}

Molecular gas dominates the total mass budget of neutral gas in the circumnuclear regions of spiral galaxies. Molecular line observations are thus required to provide a complete view of the distribution and kinematics of neutral gas, a key to understanding how and for how long activity can be sustained in active galactic nuclei (AGN). Furthermore, AGN are able to inject vast amounts of energy in the surrounding
Interstellar Medium (ISM) of their hosts, most particularly, in their circumnuclear molecular disks. Therefore, a complete understanding of the complex status of ISM requires also mm-line observations to probe the feedback influence of activity on the energy balance/redistribution of the ISM in AGN. The central gas disks of radio-loud AGN (hosted by early-type galaxies) may well be also predominantly molecular. Unfortunately, most of the molecular line observations of radio galaxies to date have been plagued by insufficient sensitivity or low spatial resolution (e.g., Evans et al.~\cite{eva05}; Prandoni et al.~\cite{pra07}; Saripalli \& Mack~\cite{sar07}). As molecular lines are expected to be in emission but also in absorption (against the strong radio continuum sources) in radio galaxies, high spatial resolution and high sensitivity observations are needed to, at the least, reduce the bias due to the confusion of absorption and emission in the beam. Moreover, the combination of emission and absorption lines is paramount to get a complete 2D picture of the controversial dynamical status of the gas disks in radio galaxies. Conclusions based {\it only} on absorption line studies can be very misleading if we want to establish if the gas kinematics are dominated by inflow motions, outflow motions or circular rotation.

In this paper we present the first results of a molecular line study of two nearby radio galaxies: the compact symmetric object (CSO) 4C~31.04 and the Fanaroff-Riley type II object (FR~II) 3C~293.  The study is based on high-resolution (0.5$\arcsec$--1$\arcsec$) observations done with the Plateau de Bure interferometer (PdBI) that make use of several species (like CO, HCO$^+$ and HCN) adapted to probe the emission and the absorption of molecular gas in these sources. A study of the content, distribution, and kinematics of cold gas via molecular lines is a key to elucidating the evolutionary link between {\it young} radio galaxies, thought to be represented by CSO, and the more {\it evolved} radio galaxy class, of which FR~I and FR~II are thought to be fair representatives.

\section{{\it Young} radio galaxies: the case of 4C~31.04}

 4C\,31.04 is a nearby CSO (z$\sim$0.06) lying at the nucleus of the giant elliptical galaxy MCG 5-8-18. HI absorption was first detected in this CSO by Van Gorkom et al.~(\cite{van89}) (see also Conway~\cite{con96}). The HST maps of Perlman et al.~(\cite{per01}) reveal that the optical image of 4C\,31.04 is permeated with dusty features; these consist of a disk and a spiral structure. The central disk is perpendicular to the axis of the $\sim$100~pc-size radio source identified in the 5\,GHz VLBA image of Giovannini et al.~(\cite{gio01}).

Garc\'{\i}a-Burillo et al.~(\cite{gb07})(hereafter GB07) have used the PdBI to probe the gas and dust content of 4C\,31.04 by observing the HCO$^+$(1--0) and $^{12}$CO(2--1) molecular lines and their underlying continuum emission with high spatial resolution ($\leq$1$\arcsec$).
The continuum maps obtained at 1mm and 3mm with the PdBI show strong emission in 4C\,31.04 (Fig.\,\ref{4c31-1}a--b). The continuum distribution at 1mm is fitted by a $\sim$40\,mJy point source centered on the AGN and a Gaussian source of $\sim$20\,mJy flux density and 1.1\,kpc$\times$0.1\,kpc deconvolved size (suggestive of a disk). The $\sim$160\,mJy continuum source detected at 3mm is point-like. The point source detected at 1mm and at 3mm is the higher frequency counterpart of the non-thermal emission detected at cm-wave-lengths. The spectral index ($\alpha$) of the total continuum emission between 84\,GHz and 218\,GHz (with S$_{\nu}$$\sim$$\nu^{-\alpha}$) confirms that the emission is dominated by non-thermal processes: $\alpha$$\geq$1 (see Fig.~\ref{4c31-1}b). However, the extended disk detected at 1mm betrays thermal emission from dust. In particular, the 1mm disk is tightly linked to the disk identified in the  R-H color image of Perlman et al.~(\cite{per01}) (Fig.\,\ref{4c31-1}a). The dust mass budget (M$_{dust}$) of the disk of 4C\,31.04 has been derived by GB07 by fitting the fluxes measured by IRAS (at 100$\mu$m and 60$\mu$m) and the PdBI (at 1mm). The best fit gives a value for M$_{dust}$ of (4--7)$\times$10$^{8}$M$_{\odot}$. With a standard molecular gas/dust ratio~$\sim$100, GB07 estimate that the total molecular gas mass of the disk is M$_{gas}$$\sim$(4--7)$\times$10$^{10}$M$_{\odot}$.

The existence of a massive gas reservoir in 4C\,31.04 is confirmed by the detection of significantly strong HCO$^{+}$(1--0) emission coming from a rotating molecular disk (Fig.\,\ref{4c31-1}c). 
The HCO$^{+}$(1--0) disk, of$\sim$1.4\,kpc-size, is detected both in emission and absorption. 
As shown in Fig.\,\ref{4c31-1}c and  Fig.\,\ref{4c31-2}, HCO$^{+}$(1--0) is detected in emission over $\sim$950kms$^{-1}$ in two separate velocity intervals: channels {\bf I} and {\bf III} (as defined in Fig.\,\ref{4c31-2}). The strongest emission ({\bf III}) peaks NE of the AGN (offset A in Fig.\,\ref{4c31-1}c). Emission in {\bf I} peaks SW of the AGN (offset B in Fig.\,\ref{4c31-1}c). In  contrast, the line is detected in absorption over $\sim$250kms$^{-1}$ against the continuum source of the AGN in channel {\bf II} (as defined in  Fig.\,\ref{4c31-2}). Assuming that the total width of the HCO$^+$ emission profile is twice the projected rotational velocity ($v_{rot}$), GB07 derive $v_{rot} \sim$500kms$^{-1}$ for an edge-on disk geometry and a value for the systemic velocity (v$_{sys}$) that corresponds to $z$=0.0602. The gas seen in absorption is globally blue-shifted by $\sim$150\,km\,s$^{-1}$ with respect to the value of v$_{sys}$ derived from the HCO$^{+}$ emission profile. This result indicates that the gas causing the absorption is subject to strong non-circular motions. Interestingly, GB07 report on further evidences of morphological distortions identified in the disk. The SW side of the molecular disk is offset from the dust disk, a result suggestive of a tilted or warped distribution for the gas.

The HCO$^{+}$ line emission is a good tracer of the dense molecular gas content (n(H$_2$)$>$10$^{4}$cm$^{-3}$). GB07 derive that the mass of the dense molecular gas in 4C\,31.04 amounts to $\sim$4$\times$10$^9$M$_{\odot}$. In particular, GB07 estimate that the {\it total} molecular gas mass would range between~$\sim$5$\times$10$^9$M$_{\odot}$ and~$\sim$4$\times$10$^{10}$M$_{\odot}$, a result consistent with the value derived from the continuum emission of dust. Most remarkably, the molecular gas masses derived from mm line and continuum observations of 4C\,31.04 are one-to-two orders of magnitude higher than those inferred from HI absorption line studies or from optical extinction measurements in this CSO (Conway~\cite{con96}; Perlman et al~\cite{per01}). As such, this flagrant disagreement highlights the need of using mm interferometers to unbiasedly probe the total gas content of radio galaxies. 

\section{{\it Evolved} radio galaxies: the case of 3C~293}

3C\,293 is a nearby ($z$$\sim$0.045) moderately large ($\sim$200~kpc) FR~II radio galaxy located at the nucleus of the elliptical vv5-33-12. The inner $\sim$4~kpc structure of its radio source has been resolved into a jet feature that runs mostly east-west (e.g., Beswick et al.~\cite{bes04}). Extensive dust lanes are identified in the HST images (Martel et al.~\cite{mar99}). The moderate ($\sim$3$\arcsec$) resolution CO interferometer map of Evans et al.~(\cite{eva99}) shows that molecular gas is distributed in a hardly resolved disk around the AGN. Using the WSRT array, Morganti et al.~(\cite{mor03}) report the detection of a broad ($\sim$1400 kms$^{-1}$) HI absorption feature. The absorption profile identifies a surprisingly fast outflow of neutral gas in 3C~293. 

\begin{figure}
\begin{center}
\includegraphics[width=5cm, angle=-90]{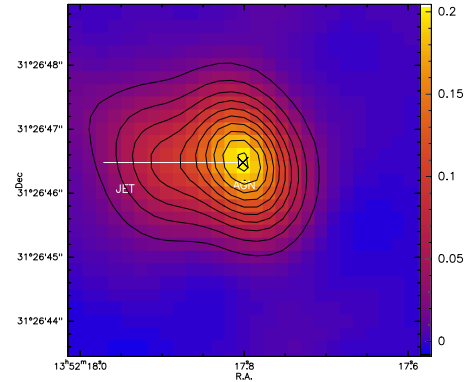}
\caption{The 3mm-continuum PdBI map of 3C\,293 (contours: 20 to 200 in steps of 20~mJy), 
showing the AGN source and the jet component to the east. See GB08 for details.}
\label{3c293-1}
\end{center}
\end{figure}

\begin{figure}
\begin{center}
\includegraphics[width=6.5cm, angle=-90]{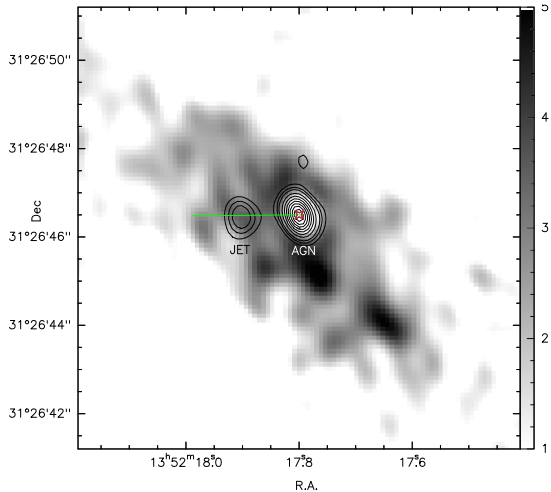}
\caption{We overlay the 1mm-continuum PdBI map of 3C\,293 (contour levels: 10, 15, 20 to 90 in steps of 10$\%$ of the peak absorption $\sim$--110~mJy) with the $^{12}$CO(2-1) integrated intensity emission map (grey scale shown in Jy~km~s$^{-1}$ units). The eastern component of the jet
is also detected at 1mm. See GB08 for details.}
\label{3c293-2}
\end{center}
\end{figure}

We have recently used the PdBI to probe the gas content of 3C~293, observing the HCO$^+$(1--0), HCN$^+$(1--0) and $^{12}$CO(2--1) lines and their continuum emission at~$\leq$1$\arcsec$ spatial resolution (Garc\'{\i}a-Burillo et al.~\cite{gb08}, hereafter GB08). We have detected the continuum emission at 3mm and 1mm not only from the AGN core, but also from the hot spot component of the inner jet, located $\sim$1.2$\arcsec$ east of the core (Figs.~\ref{3c293-1} and \ref{3c293-2}). This hot spot coincides with the continuum source {\it 'E1'} detected at cm wavelenghts by Beswick et al.~(\cite{bes04}). The molecular gas disk feeding the AGN has been mapped and fully resolved in the CO(2--1) line, seen in emission everywhere in the disk but towards the AGN source (Fig.~\ref{3c293-2}). The disk, of $\sim$7~kpc diameter and significantly high mass content ($\sim$5$\times$10$^9$M$_{\odot}$), is roughly symmetric and displays a very regular rotating pattern around the AGN (see GB08 for a detailed description).

\begin{figure}
\begin{center}
\includegraphics[width=6.0cm, angle=-90]{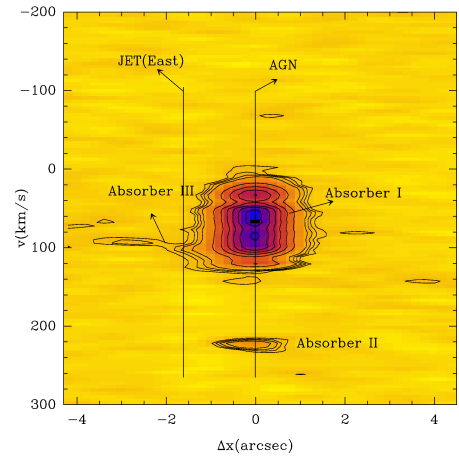}
\caption{We show in this figure the HCO$^{+}$(1--0) p-v plot taken along the apparent axis of the jet at 3mm (roughly oriented east-west). We highlight the presence of a narrow ($<$30kms$^{-1}$) absorption system on the east side of the jet (absorber III). There is also a narrow absorption line at v=250kms$^{-1}$ that betrays absorption against the AGN, superposed on a much deeper and broader absorption detected around v=60kms$^{-1}$. Levels are 4, 6, 8, 10, 20 to 80 in steps of 15$\%$ of the peak absorption $\sim$--180~mJy. See GB08 for details.}
\label{3c293-3}
\end{center}
\end{figure}

 We have also detected a complex system of HCO$^{+}$ and HCN lines seen in absorption against the AGN mm-core and against the eastern hot spot component of the radio jet (Fig.~\ref{3c293-3}). There is a narrow absorption line towards the AGN (absorber II), superposed on a much deeper and significantly broader absorption profile (absorber I), as shown in Fig.~\ref{3c293-3}. We have also detected a very narrow ($<$20kms$^{-1}$) absorption system towards the hot spot component (called absorber III in Fig.~\ref{3c293-3}). These narrow absorption lines are likely produced by molecular clouds which are located at large radii in the disk. In this scenario, the clouds are passing in front of the jet, but they are seemingly not interacting with the radio plasma. In this context, it is worth noting that, down to the limit of our sensitivity, we have not detected the molecular counterpart of the fast HI outflow.

\section{Discussion}

The mm line and continuum observations of 4C\,31.04 and 3C~293 have revealed the existence of significant amounts of molecular gas in the two radio galaxies. The derived gas masses are comparable to those identified in ULIRG (San-ders et al.~\cite{san91}). 
Although of similar gas mass, the molecular disks of 4C\,31.04 and 3C~293 
show significantly different properties. While drawing firm conclusions is still premature, it is tempting to link these differences with the fact that these two sources are thought to represent two different evolutionary stages of the radio-loud phenomenon.

The comparatively smaller size of the molecular disk of 4C\,31.04, implies that the surface density is an order of magnitude higher in this CSO compared to 3C~293. In this context, it is worth noting that the CO(2--1) emission is barely detected in the PdBI map of 4C~31.04. This is in clear constrast with the 
result obtained by GB08 in 3C~293. The previous PdBI observations at 1mm published by GB07 suggest that the HCO$^+$(1--0)/CO(2--1) is surprisingly large in 4C~31.04. Assuming a CO(2--1)/CO(1--0) brightness temperature ratio of 1, we infer that the  HCO$^+$(1--0)/CO(1--0) ratio in 4C~31.04 would be $\sim$0.3, nearly an order of magnitude larger than the ratio measured in 3C~293. This indicates that the fraction of dense molecular gas is extremely high in 4C~31.04 compared to a normal galaxy, and seemingly larger than that estimated for 3C~293. This is consistent with the comparatively higher gas surface density measured in 4C\,31.04, as derived from the molecular disk sizes of these radio galaxies.

The analytical models of Carvalho~(\cite{car98}) predict that a clumpy medium with M$_{gas}$$\sim$10$^9$--10$^{10}$M$_{\odot}$ is able to confine a radio source on the scale of 0.5--1\,kpc. In the case of 4C\,31.04, the bulk of M$_{gas}$ is seen to be in a disk perpendicular to the radio source's axis, a geometry that would decrease the efficiency of frustration, however. Within the frustration scheme, it is foreseen that a disturbed ISM would reflect the interaction between the radio plasma and the den-se confining cocoon. In the case of 4C\,31.04, there is evidence that a radio lobe-ISM interaction may be at work. The HST image of Perlman et al~(\cite{per01}) shows cone-like features aligned with the radio axis of the source (see Fig.~\ref{4c31-1}c), the likely signature of gas shocked by the jet. Moreover, the distribution and kinematics of molecular gas probed by the HCO$^+$ observations discussed by GB07 illustrate that the detected rotating disk is not in a fully relaxed state. The revealed distortions indicate that the disk is still settling after a merger or an event of gas accretion. Alternatively, the jet and the cone-like features may be interacting with the disk, thus producing the reported distortions. In the case of 3C~293, the distribution and kinematics of molecular gas are reminiscent of a regularly rotating and dynamically relaxed disk. Contrary to the case of 4C\,31.04, there are few signatures, if any, of an ongoing interaction between the jet and the molecular gas disk in 3C~293.


\end{document}